\documentclass[11pt]{revtex4}
\usepackage{amssymb,epsf}
\usepackage{latexsym}
\begin{document}

\title{Thermodynamics of rotating solutions in
$(n+1)$-dimensional Einstein-Maxwell-dilaton gravity}
\author{A. Sheykhi$^{1}$, M. H. Dehghani$^{1,2}$
\footnote{email: mhd@shirazu.ac.ir}, N. Riazi$^{1}$
\footnote{email: riazi@physics.susc.ac.ir} and J. Pakravan$^{1}$}
\address{$1$. Physics Department and Biruni Observatory, Shiraz University, Shiraz 71454, Iran\\
         $2$. Research Institute for Astrophysics and Astronomy of Maragha (RIAAM), Maragha, Iran}

\begin{abstract}
We construct a class of charged, rotating solutions of
$(n+1)$-dimensional Einstein-Maxwell-dilaton gravity with
Liouville-type potentials and investigate their properties. These
solutions are neither asymptotically flat nor (anti)-de Sitter. We
find that these solutions can represent black brane, with two
inner and outer event horizons, an extreme black brane or a naked
singularity provided the parameters of the solutions are chosen
suitably. We also compute temperature, entropy, charge, electric
potential, mass and angular momentum of the black brane solutions,
and find that these quantities  satisfy the first law of
thermodynamics. We find a Smarr-type formula and perform a
stability analysis by computing the heat capacity in the canonical
ensemble. We find that the system is thermally stable for $\alpha
\leq 1$, while for $\alpha > 1$ the system has an unstable phase.
This is incommensurate with the fact that there is no Hawking-Page
phase transition for black objects with zero curvature horizon.
\end{abstract}

\maketitle

\section{Introduction\label{Intr}}

It seems likely that gravity is not given by the Einstein action,
at least at sufficiently high energy scales. The most promising
alternative seems to be that offered by string theory, where the
gravity becomes scalar-tensor in nature. In the low energy limit
of the string theory, one recovers Einstein gravity along with a
scalar dilaton field which is non-minimally coupled to gravity
\cite{Wit1}. When a dilaton is coupled to Einstein-Maxwell theory,
it has profound consequences for the black hole solutions. This
fact may be seen in the case of rotating Einstein-Maxwell-dilaton
(EMD) black holes of Kaluza-Klein theory with coupling constant
$\alpha =\sqrt{3}$ which does not possess the gyromagnetic ratio
$g=2$ of Kerr-Newman black hole \cite{Hor1}. Thus, it is worth to
find exact solutions of EMD gravity for an arbitrary coupling
constant, and investigate how the properties of black holes/branes
are modified when a dilaton is present. Also it is interesting to
investigate the thermal stability of black holes/branes in the
presence of dilaton, and find out the effect of dilaton in the
stability of the solutions.

Exact charged dilaton black hole solutions of EMD gravity in the
absence of a dilaton potential have been constructed by many
authors \cite{CDB1, CDB2}. The dilaton changes the casual
structure of the spacetime and leads to curvature singularities at
finite radii. These black holes are asymptotically flat. In recent
years, non-asymptotically flat black hole spacetimes are
attracting much interest in connection with the so called AdS/CFT
correspondence. Black hole spacetimes which are neither
asymptotically flat nor (anti)-de Sitter [(A)dS] have been found
and investigated by many authors. The uncharged solutions have
been found in \cite{MW}, while the charged solutions have been
considered in \cite{PW}. In the presence of Liouville-type
potential, static charged black hole solutions have been
discovered with positive, zero or negative constant curvature
horizons \cite{CHM,Cai}. Recently, the properties of these black
hole solutions which are not asymptotically AdS or dS, have been
studied \cite{Clem}. These exact solutions are all static. Charged
rotating dilaton black holes with curved horizon have not been
constructed in four or higher dimensions for an arbitrary coupling
constant and arbitrary rotation parameter. Indeed, exact magnetic
rotating solutions have been considered in three dimensions
\cite{Dia}, while exact rotating black hole solutions in four
dimensions have been obtained only for some limited values of the
coupling constant \cite{Fr}. For general dilaton coupling, the
properties of rotating charged dilaton black holes only with
infinitesimally small charge \cite{Cas} or small angular momentum
in four \cite{Hor2}  and five \cite{SR} dimensions have been
investigated. For arbitrary values of angular momentum and charge
only a numerical investigation has been done \cite{Klei}.
Recently, two classes of magnetic rotating solutions in
four-dimensional Einstein-Maxwell-dilaton gravity with
Liouville-type potential have been constructed by one of us \cite
{Deh1}. Although these solutions are not black holes and represent
spacetimes with conic singularities, charged black string
solutions in four-dimensional EMD gravity have also been
constructed \cite{Deh2}. Till now, exact rotating charged dilaton
black hole/brane solutions for an arbitrary coupling constant in
more than four dimensions have not been constructed. Our aim in
this paper is to construct exact, rotating
charged dilaton black brane in $%
(n+1)$ dimensions for an arbitrary value of coupling constant and
investigate their properties. Specially, we want to perform a
stability analysis and investigate the effect of dilaton on the
stability of the solutions.

The outline of the paper is as follows. Section \ref{Field} is
devoted to a brief review of the field equations and general
formalism of calculating the
conserved quantities. In Sec. \ref{Charged}, we construct the $(n+1)$%
-dimensional charged rotating dilaton black branes with a complete set of
rotation parameters and investigate their properties. In Sec. \ref{Therm},
we obtain the conserved and thermodynamical quantities of the $(n+1)$%
-dimensional black brane solutions and show that these quantities
satisfy the first law of thermodynamics. In Sec. \ref{stab}, we
perform a stability analysis and show that the dilaton creates an
unstable phase for the solutions. We finish our paper with some
concluding remarks.

\section{General Formalism} \label{Field}

The action of $(n+1)$-dimensional dilaton Einstein-Maxwell gravity
with one scalar field can be written as \cite{CHM}\newline
\begin{eqnarray}
I_{G} &=&-\frac{1}{16\pi }\int_{\mathcal{M}}d^{n+1}x\sqrt{-g}\left( \mathcal{%
R}\text{ }-\frac{4}{n-1}(\nabla \Phi )^{2}-V(\Phi )-e^{-4\alpha \Phi
/(n-1)}F_{\mu \nu }F^{\mu \nu }\right)   \nonumber \\
&&+\frac{1}{8\pi }\int_{\partial \mathcal{M}}d^{n}x\sqrt{-\gamma
}\Theta (\gamma ),  \label{Act}
\end{eqnarray}
where $\mathcal{R}$ is the Ricci scalar curvature, $\Phi $ is the
dilaton field and $V(\Phi )$ is a potential for $\Phi $. $\alpha $
is a constant determining the strength of coupling of the scalar
and electromagnetic field, $F_{\mu \nu }=\partial _{\mu }A_{\nu
}-\partial _{\nu }A_{\mu }$ is the electromagnetic  field tensor
and $A_{\mu }$ is the electromagnetic potential. The last term in
Eq. (\ref{Act}) is the Gibbons-Hawking boundary term which is
chosen such that the variational principle is well-defined. The manifold $%
\mathcal{M}$ has metric $g_{\mu \nu }$ and covariant derivative $\nabla
_{\mu }$. $\Theta $ is the trace of the extrinsic curvature $\Theta ^{ab}$
of any boundary(ies) $\partial \mathcal{M}$ of the manifold $\mathcal{M}$,
with induced metric(s) $\gamma _{ab}$. In this paper, we consider the action
(\ref{Act}) with a Liouville type potential,
\begin{equation}
V(\Phi )=2\Lambda e^{4\alpha \Phi /(n-1)},  \label{v1}
\end{equation}
where $\Lambda $ is a constant which may be referred to as the
cosmological constant, since in the absence of the dilaton field
($\Phi =0$) the action (\ref{Act}) reduces to the action of
Einstein-Maxwell gravity with cosmological constant. The equations
of motion can be obtained by varying the action (\ref{Act}) with
respect to the gravitational field $g_{\mu \nu }$, the dilaton
field $\Phi $ and the gauge field $A_{\mu }$ which yields the
following field equations
\begin{equation}
\mathcal{R}_{\mu \nu }=\frac{4}{n-1}\left( \partial _{\mu }\Phi \partial
_{\nu }\Phi +\frac{1}{4}g_{\mu \nu }V(\Phi )\right) +2e^{\frac{-4\alpha \Phi
}{n-1}}\left( F_{\mu \eta }F_{\nu }^{\text{ }\eta }-\frac{1}{2(n-1)}g_{\mu
\nu }F_{\lambda \eta }F^{\lambda \eta }\right) ,  \label{FE1}
\end{equation}
\begin{equation}
\nabla ^{2}\Phi =\frac{n-1}{8}\frac{\partial V}{\partial \Phi }-\frac{\alpha
}{2}e^{-\frac{4\alpha \Phi }{n-1}}F_{\lambda \eta }F^{\lambda \eta },
\label{FE2}
\end{equation}
\begin{equation}
\partial _{\mu }\left( \sqrt{-g}e^{\frac{-4\alpha \Phi }{n-1}}F^{\mu \nu
}\right) =0,  \label{FE3}
\end{equation}
The conserved charges of the spacetime can be calculated through
the use of the substraction method of Brown and York \cite{BY}.
Such a procedure causes the resulting physical quantities to
depend on the choice of reference background. For asymptotically
(A)dS solutions, the way that one deals with these divergences is
through the use of counterterm method inspired by (A)dS/CFT
correspondence \cite{Mal}. However, in the presence of a
non-trivial dilaton field, the spacetime may not behave as either dS ($%
\Lambda >0$) or AdS ($\Lambda <0$). In fact, it has been shown that with the
exception of a pure cosmological constant potential, where $\alpha =0$, no
AdS or dS static spherically symmetric solution exist for Liouville-type
potential \cite{PW}. But, as in the case of asymptotically AdS spacetimes,
according to the domain-wall/QFT (quantum field theory) correspondence \cite
{Sken}, there may be a suitable counterterm for the stress energy tensor
which removes the divergences. In this paper, we deal with the spacetimes
with zero curvature boundary [$R_{abcd}(\gamma )=0$], and therefore the
counterterm for the stress energy tensor should be proportional to $\gamma
^{ab}$. Thus, the finite stress-energy tensor in $(n+1)$-dimensional
Einstein-dilaton gravity with Liouville-type potential may be written as
\begin{equation}
T^{ab}=\frac{1}{8\pi }\left[ \Theta ^{ab}-\Theta \gamma ^{ab}+\frac{n-1}{l_{%
\mathrm{eff}}}\gamma ^{ab}\right] ,  \label{Stres}
\end{equation}
where $l_{\mathrm{eff}}$ is given by
\begin{equation}
l_{\mathrm{eff}}^{2}=\frac{(n-1)(\alpha ^{2}-n)}{2\Lambda }e^{\frac{-4\alpha
\Phi }{n-1}}.  \label{leff}
\end{equation}
In the particular case $\alpha =0$, the effective
$l_{\mathrm{eff}}^{2}$ of Eq. (\ref{leff}) reduces to
$l^{2}=-n(n-1)/2\Lambda $ of the AdS spacetimes. The first two
terms in Eq. (\ref{Stres}) is the variation of the action
(\ref{Act}) with respect to $\gamma _{ab}$, and the last term is
the counterterm which removes the divergences. One may note that
the counterterm has the same form as in the case of asymptotically
AdS solutions with zero curvature boundary, where $l$ is replaced
by $l_{\mathrm{eff}}$. To compute the
conserved charges of the spacetime, one should choose a spacelike surface $%
\mathcal{B}$ in $\partial \mathcal{M}$ with metric $\sigma _{ij}$,
and write the boundary metric in ADM (Arnowitt-Deser-Misner) form:
\[
\gamma _{ab}dx^{a}dx^{a}=-N^{2}dt^{2}+\sigma _{ij}\left( d\varphi
^{i}+V^{i}dt\right) \left( d\varphi ^{j}+V^{j}dt\right) ,
\]
where the coordinates $\varphi ^{i}$ are the angular variables
parameterizing the hypersurface of constant $r$ around the origin, and $N$
and $V^{i}$ are the lapse and shift functions respectively. When there is a
Killing vector field $\mathcal{\xi }$ on the boundary, then the quasilocal
conserved quantities associated with the stress tensors of Eq. (\ref{Stres})
can be written as
\begin{equation}
Q(\mathcal{\xi )}=\int_{\mathcal{B}}d^{n-1}\varphi \sqrt{\sigma }T_{ab}n^{a}%
\mathcal{\xi }^{b},  \label{charge}
\end{equation}
where $\sigma $ is the determinant of the metric $\sigma _{ij}$, $\mathcal{%
\xi }$ and $n^{a}$ are the Killing vector field and the unit
normal vector on the boundary $\mathcal{B}$. For boundaries with
timelike ($\xi =\partial /\partial t$) and rotational ($\varsigma
=\partial /\partial \varphi $) Killing vector fields, one obtains
the quasilocal mass and angular momentum
\begin{eqnarray}
M &=&\int_{\mathcal{B}}d^{n-1}\varphi \sqrt{\sigma }T_{ab}n^{a}\xi ^{b},
\label{Mastot} \\
J &=&\int_{\mathcal{B}}d^{n-1}\varphi \sqrt{\sigma }T_{ab}n^{a}\varsigma
^{b},  \label{Angtot}
\end{eqnarray}
provided the surface $\mathcal{B}$ contains the orbits of $\varsigma $.
These quantities are, respectively, the conserved mass and angular momenta
of the system enclosed by the boundary $\mathcal{B}$. Note that they will
both depend on the location of the boundary $\mathcal{B}$ in the spacetime,
although each is independent of the particular choice of foliation $\mathcal{%
B}$ within the surface $\partial \mathcal{M}$.\newline
\section{$(n+1)$-dimensional Solutions in EMD Gravity}\label{Charged}
\label{field}Our aim here is to construct the $(n+1)$-dimensional
rotating solutions of the field equations (\ref{FE1})-(\ref{FE3})
with $k$ rotation
parameters and investigate their properties. The rotation group in $(n+1)$%
-dimensions is $SO(n)$ and therefore the number of independent rotation
parameters for a localized object is equal to the number of Casimir
operators, which is $[n/2]\equiv k$, where $[x]$ is the integer part of $x$.
The metric of $(n+1)$-dimensional rotating solution with cylindrical or
toroidal horizons and $k$ rotation parameters can be written as \cite{awad}
\begin{eqnarray}
ds^{2} &=&-f(r)\left( \Xi dt-{{\sum_{i=1}^{k}}}a_{i}d\phi _{i}\right) ^{2}+%
\frac{r^{2}}{l^{4}}R^{2}(r){{\sum_{i=1}^{k}}}\left( a_{i}dt-\Xi l^{2}d\phi
_{i}\right) ^{2}  \nonumber  \label{metric} \\
&&-\frac{r^{2}}{l^{2}}R^{2}(r){\sum_{i<j}^{k}}(a_{i}d\phi _{j}-a_{j}d\phi
_{i})^{2}+\frac{dr^{2}}{f(r)}+\frac{r^{2}}{l^{2}}R^{2}(r)dX^{2},  \nonumber \\
\Xi ^{2} &=&1+\sum_{i=1}^{k}\frac{a_{i}^{2}}{l^{2}},  \label{Met3}
\end{eqnarray}
where $a_{i}$'s are $k$ rotation parameters. The functions $f(r)$ and $R(r)$
should be determined and $l$ has the dimension of length which is related to
the cosmological constant $\Lambda $ for the case of Liouville-type
potential with constant $\Phi $. The angular coordinates are in the range $%
0\leq \phi _{i}\leq 2\pi $ and $dX^{2}$ is the Euclidean metric on the $%
(n-k-1)$-dimensional submanifold with volume $\Sigma _{n-k-1}$.

The Maxwell equation (\ref{FE3}) can be integrated immediately to
give
\begin{eqnarray}
F_{tr} &=&\frac{q\Xi e^{\frac{4\alpha \Phi }{n-1}}}{(rR)^{n-1}}
\nonumber
\label{Ftr} \\
F_{\phi _{i}r} &=&-\frac{a_{i}}{\Xi }F_{tr}.
\end{eqnarray}
where $q$, is an integration constant related to the electric
charge of the brane. In order to solve the system of equations
(\ref{FE1}) and (\ref{FE2}) for three unknown functions $f(r)$,
$R(r)$ and $\Phi (r)$, we make the ansatz
\begin{equation}
R(r)=e^{2\alpha \Phi /(n-1)}  \label{Rphi}
\end{equation}
Using (\ref{Rphi}), the Maxwell fields (\ref{Ftr}) and the metric (\ref{Met3}%
), one can easily show that equations (\ref{FE1}) and (\ref{FE2})
have solutions of the form
\begin{equation}
f(r)=\frac{2\Lambda (\alpha ^{2}+1)^{2}b^{2\gamma }}{(n-1)(\alpha ^{2}-n)}%
r^{2(1-\gamma )}-\frac{m}{r^{(n-1)(1-\gamma )-1}}+\frac{2q^{2}(\alpha
^{2}+1)^{2}b^{-2(n-2)\gamma }}{(n-1)(\alpha ^{2}+n-2)}r^{2(n-2)(\gamma -1)},
\label{f}
\end{equation}
\begin{equation}
\Phi (r)=\frac{(n-1)\alpha }{2(1+\alpha ^{2})}\ln (\frac{b}{r}),  \label{phi}
\end{equation}
where $b$ and $m$ are arbitrary constants and $\gamma =\alpha
^{2}/(\alpha ^{2}+1)$. One may note that in the particular case
$n=3$ these solutions reduce to the four-dimensional charged
rotating dilaton black strings presented in \cite{Deh2}. In the
absence of a non-trivial dilaton ($\alpha =0 $), the above
solutions reduce to the asymptotically AdS charged rotating black
branes presented in \cite{Deh3,awad}.
\subsection*{Properties of the solutions}
In order to study the general structure of these solutions, we
first look for the curvature singularities in the presence of
dilaton gravity. It is easy to show that the Kretschmann scalar
$R_{\mu \nu \lambda \kappa }R^{\mu \nu \lambda \kappa }$ diverges
at $r=0$, it is finite for $r\neq 0$ and goes to zero as
$r\rightarrow \infty $. Thus, there is an essential singularity
located at $r=0$. Also, it is notable to mention that the Ricci
scaler is finite every where except at $r=0$, and goes to zero as
$r\rightarrow \infty $. The spacetime is neither asymptotically
flat nor (A)dS. As in the case of rotating black hole solutions of
the Einstein gravity, the above metric given by (\ref{Met3}) and
(\ref{f}) has both Killing and event horizons. The Killing horizon
is a null surface whose null generators are tangent to a Killing
field. It is easy to see that the Killing vector
\begin{equation}
\chi =\partial _{t}+{{{\sum_{i=1}^{k}}}}\Omega _{i}\partial _{\phi _{i}},
\label{chi}
\end{equation}
is the null generator of the event horizon, where $k$ denote the number of
rotation parameters \cite{Deh4}. As one can see from Eq. (\ref{f}), the
solution is ill-defined for $\alpha =\sqrt{n}$ with a Liouville potential ($%
\Lambda \neq 0$). The cases with $\alpha >\sqrt{n}$ and $\alpha
<\sqrt{n}$ should be considered separately.

In the first case where $\alpha >\sqrt{n}$, as $r$ goes to
infinity the dominant term is the second term, and therefore the
spacetime has a cosmological horizon for positive values of the
mass parameter, despite the
sign of the cosmological constant $\Lambda $. In the second case where $%
\alpha <\sqrt{n}$, as $r$ goes to infinity the dominant term is the first
term, and therefore there exist a cosmological horizon for $\Lambda >0$,
while there is no cosmological horizons if $\Lambda <0$ . Indeed, in the
latter case ($\alpha <\sqrt{n}$ and $\Lambda <0$) the spacetimes associated
with the solution (\ref{f}) exhibit a variety of possible casual structures
depending on the values of the metric parameters $\alpha $, $m$, $q$, and $%
\Lambda $. One can obtain the casual structure by finding the roots of $%
f(r)=0$. Unfortunately, because of the nature of the exponents in
(\ref{f}), it is not possible to find explicitly the location of
horizons for an arbitrary value of $\alpha $. But, we can obtain
some information by considering the temperature of the horizons.

One can obtain the temperature and angular velocity of the horizon
by analytic continuation of the metric. The analytical
continuation of the Lorentzian metric by $t\rightarrow i\tau $
and $a\rightarrow ia$ yields the Euclidean section, whose regularity at $%
r=r_{+}$ requires that we should identify $\tau \sim \tau +\beta _{+}$ and $%
\phi _{i}\sim \phi _{i}+\beta _{+}\Omega _{i}$, where $\beta _{+}$ and $%
\Omega _{i}$ 's are the inverse Hawking temperature and the $i$th component
of angular velocity of the horizon. It is a matter of calculation to show
that
\begin{eqnarray}
T_{+} &=&\frac{f^{\text{ }^{\prime }}(r_{+})}{4\pi \Xi }=\frac{1}{4\pi \Xi }%
\left( \frac{(n-\alpha ^{2})m}{\alpha ^{2}+1}{r_{+}}^{(n-1)(\gamma -1)}-%
\frac{4q^{2}(\alpha ^{2}+1)b^{-2(n-2)\gamma }}{(\alpha ^{2}+n-2){r_{+}}%
^{\gamma }}{r_{+}}^{(2n-3)(\gamma -1)}\right)  \nonumber \\
&=&-\frac{2(1+\alpha ^{2})}{4\pi \Xi (n-1)}\left( \Lambda b^{2\gamma
}r_{+}^{1-2\gamma }+\frac{q^{2}b^{-2(n-2)\gamma }}{{r_{+}}^{\gamma }}{r_{+}}%
^{(2n-3)(\gamma -1)}\right) ,  \label{Tem} \\
\Omega _{i} &=&\frac{a_{i}}{\Xi l^{2}}.  \label{Om1}
\end{eqnarray}
Equation (\ref{Tem}) shows that the temperature is negative for the two
cases of (\emph{i}) $\alpha >\sqrt{n}$ despite the sign of $\Lambda $, and (%
\emph{ii}) positive $\Lambda $ despite the value of $\alpha $. As
we argued above in these two cases we encounter with cosmological
horizons, and therefore the cosmological horizons have negative
temperature. Numerical calculations show that the temperature of
the event horizon goes to zero as the black brane approaches the
extreme case. Thus, one can see from Eq. (\ref{Tem}) that there
exists extreme black brane only for negative $\Lambda $ and
$\alpha <\sqrt{n}$, if
\begin{equation}
r_{\mathrm{ext}}^{(3-n)\gamma +n-2}=\frac{4q^{2}(1+\alpha
^{2})^{2}b^{-2(n-2)\gamma }}{m_{\mathrm{ext}}(n-\alpha ^{2})(\alpha ^{2}+n-2)%
}
\end{equation}
where $m_{\mathrm{ext}}$ is the extremal mass parameter of black brane. If
one substitutes this $r_{\mathrm{ext}}$ into the equation $f(r_{ext})=0$,
then one obtains the condition for extreme black brane as:
\begin{equation}
m_{\mathrm{ext}}=\frac{4q^{2}(1+\alpha ^{2})^{2}b^{-2(n-2)\gamma }}{%
(n-\alpha ^{2})(\alpha ^{2}+n-2)}\left( \frac{-\Lambda b^{2\gamma (n-1)}}{%
q_{ext}^{2}}\right) ^{\frac{(3-n)\gamma +n-2}{2(\gamma -1)(1-n)}}
\label{Mext}
\end{equation}
Indeed, the metric of Eqs. (\ref{Met3}) and (\ref{f}) has two inner and
outer horizons located at $r_{-}$ and $r_{+}$, provided the mass parameter $%
m $ is greater than $m_{\mathrm{ext}}$, an extreme black brane in
the case of $m=m_{\mathrm{ext}}$, and a naked singularity if
$m<m_{\mathrm{ext}}$. Note that for $n=3$, Eq. (\ref{Mext})
reduces to the extremal mass obtained in Ref. \cite{Deh2} and  in
the absence of dilaton field ($\alpha=\gamma=0$) it reduces to
that obtained in \cite{Deh3}.
%%%%%%%%%%%%%%%%%%%%%%%%%%%%%%%%%%%%%%%%%%%%%%%%%%%%%%%%%%%%%%%%%%%%%%%%
\section{Thermodynamics \label{Therm}of black branes}
Denoting the volume of the hypersurface boundary at constant $t$ and $r$ by $%
V_{n-1}=(2\pi )^{k}\Sigma _{n-k-1}$, the mass and angular momentum per unit
volume $V_{n-1}$ of the black branes ($\alpha <\sqrt{n}$) can be calculated
through the use of Eqs. (\ref{Mastot}) and (\ref{Angtot}). We find
\begin{equation}
{M}=\frac{b^{(n-1)\gamma }}{16\pi l^{n-2}}\left( \frac{(n-\alpha ^{2})\Xi
^{2}+\alpha ^{2}-1}{1+\alpha ^{2}}\right) m,  \label{M}
\end{equation}
\begin{equation}
J_{i}=\frac{b^{(n-1)\gamma }}{16\pi l^{n-2}}\left( \frac{n-\alpha ^{2}}{%
1+\alpha ^{2}}\right) \Xi ma_{i}.  \label{J}
\end{equation}
For $a_{i}=0$ ($\Xi =1$), the angular momentum per unit volume
vanishes, and therefore $a_{i}$'s are the rotational parameters of
the spacetime. Black hole entropy typically satisfies the so
called area law of the entropy \cite {Beck}. This near universal
law applies to almost all kinds of black holes and black branes in
Einstein gravity \cite{hunt}. It is a matter of calculation to
show that the entropy per unit volume $V_{n-1}$ of the black brane
is
\begin{equation}
{S}=\frac{\Xi b^{(n-1)\gamma }r_{+}^{(n-1)(1-\gamma )}}{4l^{n-2}},
\label{Ent1}
\end{equation}
which shows that the area law holds for the black brane solutions in dilaton
gravity. Next, we calculate the electric charge of the solutions. To
determine the electric field we should consider the projections of the
electromagnetic field tensor on special hypersurfaces. The normal to such
hypersurfaces is
\begin{equation}
u^{0}=\frac{1}{N},\text{ \ }u^{r}=0,\text{ \ }u^{i}=-\frac{V^{i}}{N},
\end{equation}
where $N$ and $V^{i}$ are the lapse function and shift vector. Then the
electric field is $E^{\mu }=g^{\mu \rho }e^{\frac{-4\alpha \phi }{n-1}%
}F_{\rho \nu }u^{\nu }$, and the electric charge per unit volume $V_{n-1}$
can be found by calculating the flux of the electric field at infinity,
yielding
\begin{equation}
{Q}=\frac{\Xi q}{4\pi l^{n-2}}.  \label{chden}
\end{equation}
The electric potential $U$, measured at infinity with respect to the
horizon, is defined by \cite{Cal}
\begin{equation}
U=A_{\mu }\chi ^{\mu }\left| _{r\rightarrow \infty }-A_{\mu }\chi ^{\mu
}\right| _{r=r_{+}},
\end{equation}
where $\chi $ is the null generators of the event horizon given by
Eq. (\ref {chi}). One can easily show that the vector potential
$A_{\mu }$ corresponding to the electromagnetic tensor (\ref{Ftr})
can be written as
\begin{equation}
A_{\mu }=\frac{qb^{(3-n)\gamma }}{\Gamma r^{\Gamma }}\left( \Xi \delta _{\mu
}^{t}-a_{i}\delta _{\mu }^{i}\right) \hspace{0.5cm}{\text{(no sum on i)}}.
\label{Pot}
\end{equation}
where $\Gamma =(n-3)(1-\gamma )+1$. Therefore, the electric
potential may be obtained as
\begin{equation}
U=\frac{qb^{(3-n)\gamma }}{\Xi \Gamma {r_{+}}^{\Gamma }}.
\label{Pott}
\end{equation}
Then, we consider the first law of thermodynamics for the black
brane. In order to do this, we obtain the mass $M$ as a function
of extensive quantities $S$, $\mathbf{J}$ and $Q$. Using the
expression for the mass, the angular momenta, the entropy and the
charge given in Eqs. (\ref{M})-(\ref{chden}) and the fact that
$f(r_{+})=0$, one can obtain a Smarr- type formula as
\begin{equation}
M(S,\mathbf{J},Q)=\frac{\left( (n-\alpha ^{2})Z+\alpha ^{2}-1\right) \sqrt{%
\sum_{i}^{k}{J_{i}}^{2}}}{(n-\alpha ^{2})\sqrt{Z(Z-1)}},
\label{Msmarr}
\end{equation}
where $Z=\Xi ^{2}$ is the positive real root of the following equation:
\begin{eqnarray}
(n-1)(\alpha^2+n-2)\left(n(\alpha^2+1)Z\sqrt{Z(Z-1)}\left(\frac{4S}{Z}\right)
^{\frac{n-\alpha^2}{n-1}}-16JZ\pi
\right)\\
+32\sqrt{Z(Z-1)}\left(\frac{4S}{Z}\right)
^{\frac{n+\alpha^2-2}{1-n}}(\alpha^2+1)(n-\alpha^2)\pi^2 Q^2 =0.
\label{Zeq}
\end{eqnarray}
One may then regard the parameters $S$, $\mathbf{J}$ and $Q$ as a
complete set of extensive parameters for the mass
$M(S,\mathbf{J},Q)$ and define the intensive parameters conjugate
to $S$, $\mathbf{J}$ and $Q$. These quantities are the
temperature, the angular velocities and the electric potential
\begin{equation}\label{inte}
T=\left( \frac{\partial {M}}{\partial {S}}\right) _{\mathbf{J},
Q},\Omega _{i}=\left(
\frac{\partial {M}}{\partial {J_{i}}}\right) _{S,Q},U=\left( \frac{\partial {%
M}}{\partial {Q}}\right) _{S, \mathbf{J}}
\end{equation}
Numerical calculations show that the intensive quantities
calculated by Eq. (\ref{inte}) coincide with Eqs. (\ref{Tem}),
(\ref{Om1}) and (\ref{Pott}). Thus, these thermodynamics
quantities satisfy the first law of thermodynamics
\begin{equation}
dM=TdS+{{{\sum_{i=1}^{k}}}}\Omega _{i}d{J}_{i}+Ud{Q},
\end{equation}
\section{Stability in the canonical ensemble}\label{stab}
\begin{figure}[tbp]
\epsfxsize=10cm \centerline{\epsffile{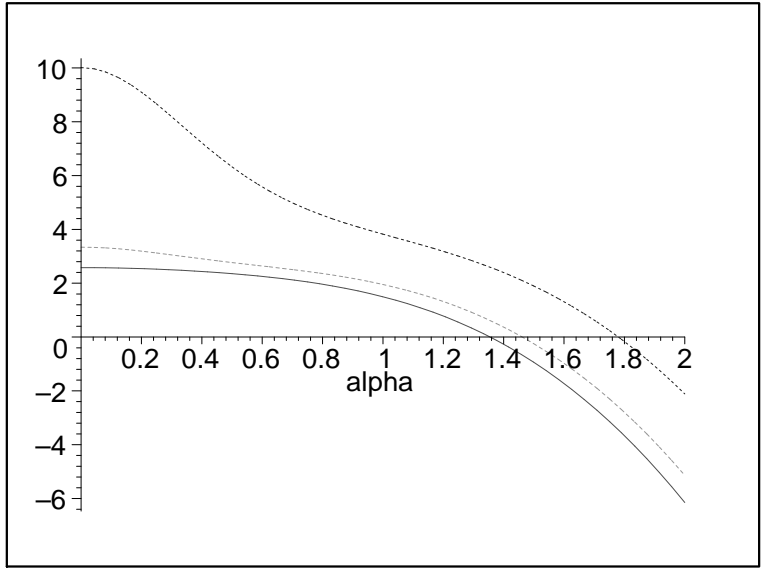}}
\caption{$(\partial ^{2}M/\partial S^{2})_{\mathbf{J},Q}$ versus
$\protect\alpha $ for $n=5$, $l=1$, $r_{+}=0.7$, $\Xi =1.25$,
$q=0$ (solid), $q=0.5$ (dotted), and $q=1$ (dashed).}
\label{Figure1}
\end{figure}

\begin{figure}[tbp]
\epsfxsize=10cm \centerline{\epsffile{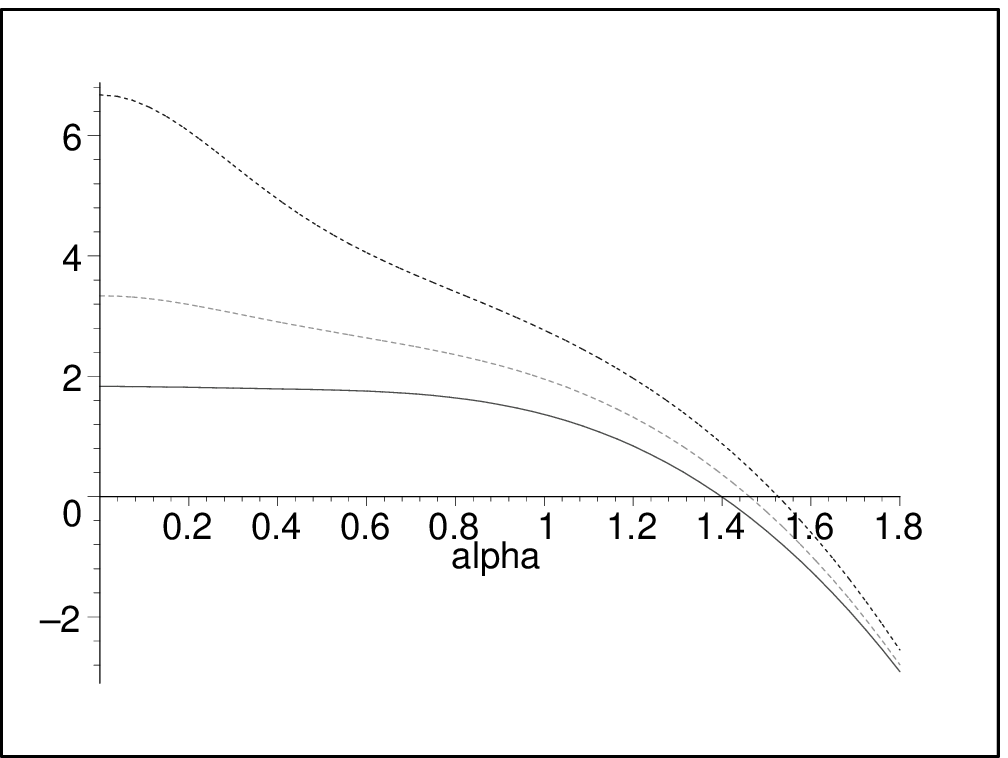}}
\caption{$(\partial ^{2}M/\partial S^{2})_{\mathbf{J},Q}$ versus
$\protect\alpha$ for $q=0.5$, $%
l=1$, $r_+=0.7$, $\Xi=1.25$, $n=4$ (solid), $n=5$ (dotted), and
$n=6$ (dashed).} \label{Figure2}
\end{figure}
\begin{figure}[tbp]
\epsfxsize=10cm \centerline{\epsffile{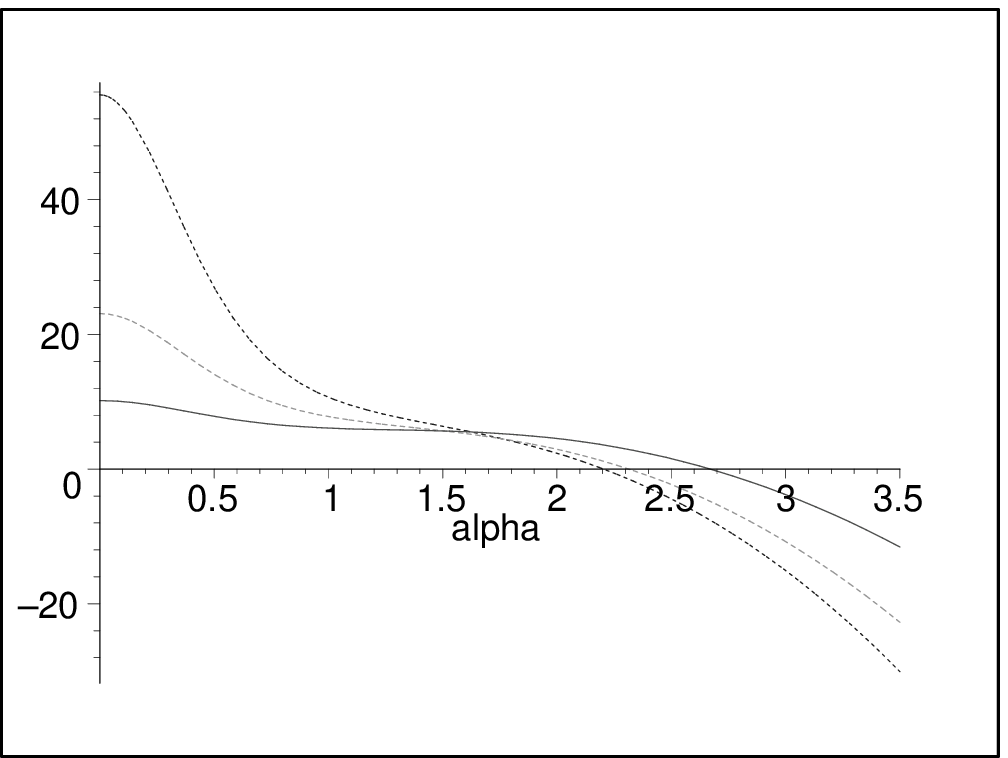}}
\caption{$(\partial ^{2}M/\partial S^{2})_{\mathbf{J},Q}$ versus
$\protect\alpha$ for $q=1.5$, $%
l=1$, $r_+=0.7$, $\Xi=1.25$, $n=4$ (solid), $n=5$ (dotted), and
$n=6$ (dashed).} \label{Figure3}
\end{figure}

\begin{figure}[tbp]
\epsfxsize=10cm \centerline{\epsffile{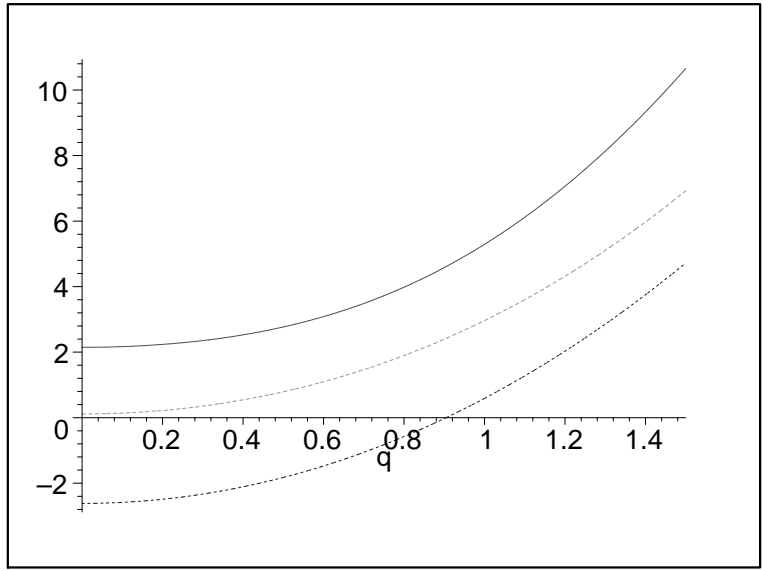}}
\caption{$(\partial ^{2}M/\partial S^{2})_{\mathbf{J},Q}$ versus $q$ for $n=6$, $l=1$, $%
r_+=0.7$, $\Xi=1.25$, $\protect\alpha=1$ (solid), $\protect\alpha=\protect%
\sqrt{2}$ (dotted), and $\protect\alpha=\protect\sqrt{3}$
(dashed).} \label{Figure4}
\end{figure}
\begin{figure}[tbp]
\epsfxsize=10cm \centerline{\epsffile{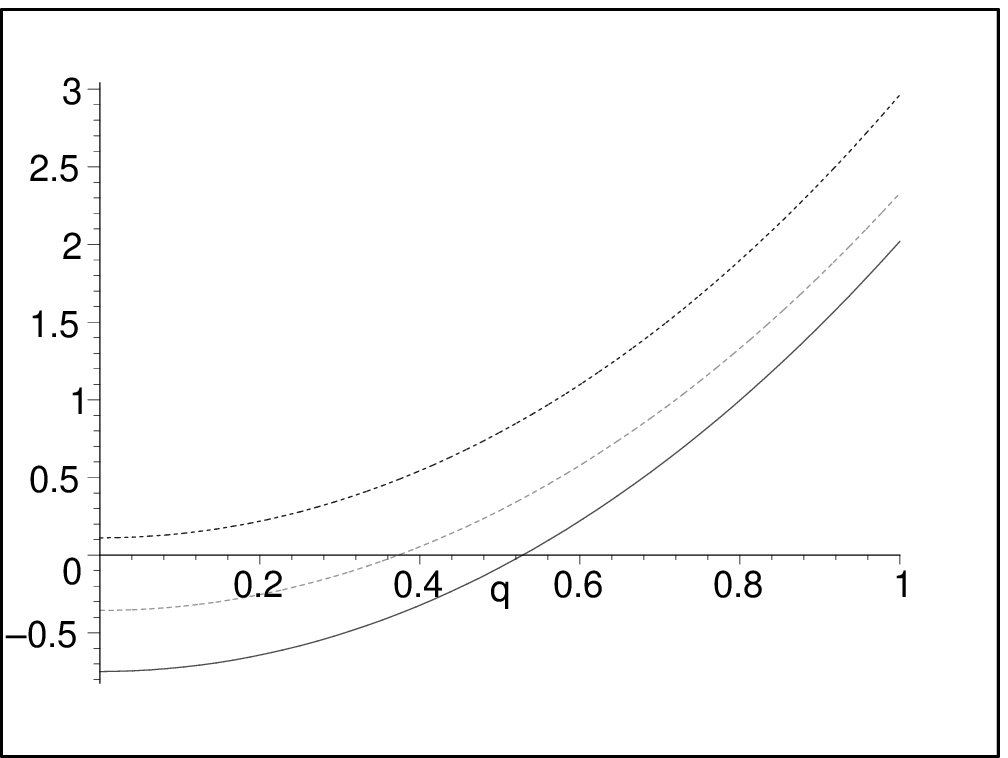}}
\caption{$(\partial ^{2}M/\partial S^{2})_{\mathbf{J},Q}$ versus
$q$
for $\protect\alpha=\protect%
\sqrt{2}$, $l=1$, $%
r_+=0.7$, $\Xi=1.25$,  $n=4$ (solid), $n=5$ (dotted), and $n=6$
(dashed).} \label{Figure5}
\end{figure}
Finally, we investigate the stability of charged rotating black
brane solutions in dilaton gravity. The stability of a
thermodynamic system with respect to small variations of the
thermodynamic coordinates is usually performed by analyzing the
behavior of the entropy $S(M,\mathbf{J},Q)$ around the
equilibrium. The local stability in any ensemble requires that
$S(M,\mathbf{J},Q)$ be a convex function of the extensive
variables or its Legendre transformation must be a concave
function of the intensive variables. The
stability can also be studied by the behavior of the energy $M(S,Q,\mathbf{J}%
)$ which should be a convex function of its extensive variable.
Thus, the local stability can in principle be carried out by
finding the determinant of the Hessian matrix of
$M(S,Q,\mathbf{J})$ with respect to its extensive variables
$X_{i}$, $\mathbf{H}_{X_{i}X_{j}}^{M}=[\partial ^{2}M/\partial
X_{i}\partial X_{j}]$ \cite{Cal, Gub}. In our case the mass $M$ is
a function of entropy, angular momenta, and charge. The number of
thermodynamic variables depends on the ensemble that is used. In
the canonical ensemble, the charge and the angular momenta are
fixed parameters, and therefore the positivity of the $(\partial
^{2}M/\partial S^{2})_{\mathbf{J},Q}$ is sufficient to ensure
local stability. Numerical calculations show that the black brane
solutions are stable independent of the value of the charge
parameter $q$ in any dimensions if $\alpha \leq 1$. Figure
\ref{Figure1} shows the behavior of the $(\partial ^{2}M/\partial
S^{2})_{\mathbf{J},Q}$\ as a function of the coupling constant
parameter $\alpha $ for different value of charge parameters $q$.
It shows  that if the dilaton coupling constant is greater than
one ($\alpha >1$), then there exists a maximum value of $\alpha
_{\mathrm{\max }}$ for which $(\partial ^{2}M/\partial
S^{2})_{\mathbf{J},Q}$ is negative provided $\alpha >\alpha
_{\mathrm{\max }}$ and positive otherwise. That is the black brane
solutions are unstable for large values of $\alpha $. It is worth
to note that $\alpha _{\mathrm{\max }}$ depends on the charge
parameter $q$  and the dimensionality of space time (see Figs.
\ref{Figure1}-\ref{Figure3}). On the other hand, figure
\ref{Figure4} shows the behavior of the $(\partial ^{2}M/\partial
S^{2})_{\mathbf{J},Q}$ as a function of the charge parameter $q$
for different value of coupling constant parameter $\alpha $. It
shows that for a fixed value of $\alpha $ there is a lower limit
for charge parameter $q$, for which our solutions are only stable
if $q>q_{\mathrm{\min }}$. In other words, they show that electric
charge makes the black brane stable for $1<\alpha<\alpha
_{\mathrm{\max }}$. This fact is the same as the case of
Reissner-Nordstr\"{o}m black hole, in which charge makes it
stable. Again, in general $q_{\mathrm{\min }}$ depends on the
dimensionality of space time for a fixed value of $\alpha $, as
one can see from Fig. \ref{Figure5}.

\section{Closing Remarks}
Unfortunately, exact rotating solutions of the Einstein equation
coupled to matter fields are difficult to find except in a limited
number of cases. Indeed, no explicit rotating charged black hole
solutions with curved
horizon have been found except for some dilaton couplings such as $\alpha =%
\sqrt{3}$ \cite{Fr} or $\alpha =1$ when the string three-form $H_{abc}$ is
included \cite{Sen}. For general dilaton coupling, the properties of charged
dilaton black holes have been investigated only for rotating solutions with
infinitesimally small angular momentum \cite{Hor2} or small charge \cite{Cas}%
. Recently, charged rotating black string solution with flat
horizon has been constructed in four dimensions. Till now, exact
rotating charged dilaton black brane solutions for an arbitrary
coupling constant in more than four
dimensions has not been constructed. In this paper, we obtained a class of $%
(n+1)$-dimensional charged rotating dilaton black brane solutions with
Liouville-type potentials . We found that these solutions are neither
asymptotically flat nor (A)dS. In the presence of Liouville-type potential,
we obtained exact solutions provided $\alpha \neq \sqrt{n}$. For $\alpha =0$%
, these solutions reduce to the charged rotating black brane
solutions of \cite{awad,Deh3}, while for $n=3$, these solutions
reduce to the four-dimensional charged rotating dilaton black
string presented in \cite
{Deh2}. We found that these solutions have a cosmological horizon for (\emph{%
i}) $\alpha >\sqrt{n}$ despite the sign of $\Lambda $, and (\emph{ii})
positive values of $\Lambda $, despite the magnitude of $\alpha $. For $%
\alpha <\sqrt{n}$, the solutions present black branes with outer and inner
horizons if $m>m_{\mathrm{ext}}$, an extreme black hole if $m=m_{\mathrm{ext}%
}$, and a naked singularity if $m<m_{\mathrm{ext}}$. The Hawking
temperature of all the above horizons were computed. We found that
the Hawking temperature is negative for inner and cosmological
horizons, and it is positive for outer horizons. We also computed
the conserved and thermodynamics quantities of the
$(n+1)$-dimensional rotating charged black brane, and found that
they satisfy the first law of thermodynamics.

We also performed a stability analysis in canonical ensemble by considering $%
(\partial ^{2}M/\partial S^{2})_{\mathbf{J},Q}$ for the charged
rotating black brane solutions in $(n+1)$ dimensions and showed
that there is no Hawking-Page phase transition in spite of charge
and angular momentum of the branes for $\alpha \leq 1$, while the
solutions have an unstable phase for large values of $\alpha $.
Indeed, for fixed values of the metric parameters, we found that
there exists a maximum value of $\alpha >1$ for which the
solutions are unstable if $\alpha >\alpha _{\mathrm{\max }}$,
where $\alpha _{\mathrm{\max }}$ depends on the dimensionality of
the spacetime and the metric parameters $m$ and $q$. This phase
behavior shows that although there is no Hawking-Page transition
for black object whose horizon is diffeomorphic to
$\mathbb{R}^{p}$ for $\alpha \leq 1$ and therefore the system is
always in the high temperature phase \cite{Wit}, but in the
presence of dilaton with $\alpha >1$ the black brane solutions
have some unstable phases. Note that the $(n+1)$-dimensional
charged rotating solutions obtained here have flat horizons. Thus,
it would be interesting if one can construct charged rotating
solutions in $(n+1)$ dimensions in the presence of dilaton and
electromagnetic fields with curved horizons.

\acknowledgments{This work has been supported in part by Research
Institute for Astrophysics and Astronomy of Maragha, Iran and also
by Shiraz University.}
%%%%%%%%%%%%%%%%%%%%%%%%%%%%%%%%%%

\end{document}